\documentclass{article}
\usepackage[utf8]{inputenc}
\usepackage[english]{babel}
\usepackage{eurosym}
\usepackage{mathrsfs}
\usepackage{graphicx}
\usepackage{fullpage}
\usepackage{caption}
\usepackage{authblk}
\usepackage{subcaption}
\usepackage{bm}
\usepackage{float}
\usepackage{amsmath,amssymb,amsfonts}
\usepackage[hyphens]{url} 
\usepackage{hyperref}
\usepackage[T1]{fontenc}
\usepackage{lmodern}
\usepackage{siunitx}
\usepackage{csquotes}
\usepackage{amsmath}{\newtheorem{simplification}{Simplification}}

\usepackage[style=ieee,doi=false,isbn=false,url=false]{biblatex}
\AtEveryBibitem{\clearlist{language}}
\AtEveryBibitem{\clearlist{note}}
\bibliography{references}

\begin{document}

\title{Probabilistic Behavioral Aggregation: A Case Study on the Nordic Power Grid}
\author[1]{Anna Büttner}
\author[1]{Frank Hellmann}
\affil[1]{Potsdam-Institute for Climate Impact Research, 14473 Potsdam, Germany}

\maketitle

   
\section*{Abstract}
    This study applies the Probabilistic Behavioral Tuning (ProBeTune) framework to transient power grid simulations to address challenges posed by increasing grid complexity. ProBeTune offers a probabilistic approach to model aggregation, using a behavioral distance measure to quantify and minimize discrepancies between a full-scale system and a simplified model. We demonstrate the effectiveness of ProBeTune on the Nordic5 (N5) test case, a model representing the Nordic power grid with complex nodal dynamics and a high share of RESs. We substantially reduce the complexity of the dynamics by tuning the system to align with a reduced swing-equation model. We confirm the validity of the swing equation with tailored controllers and parameter distributions for capturing the essential dynamics of the Nordic region. This reduction could allow interconnected systems like the Central European power grid to treat the Nordic grid as a single dynamic actor, facilitating more manageable stability assessments. The findings lay the groundwork for future research on applying ProBeTune to microgrids and other complex sub-systems, aiming to enhance scalability and accuracy in power grid modeling amidst rising complexity.
        
\section{Introduction}
    Transient simulations of power grids are at the core of every dynamic stability assessment tool. Transmission system operators employ these tools in control rooms to predict and mitigate critical states in the grid. The European Network of Transmission System Operators (ENTSO-E) has mandated that large sets of fault scenarios be simulated to improve the stability assessment of power grids. Even now, the feasibility of these approaches is limited due to the high computation times of the transient simulations. Further, as renewable energy sources (RESs) replace synchronous generation, the number of actors in the grid increases as each RES typically produces less energy than a traditional power plant. This alters the grid dynamics and increases its complexity due to the higher number of actors. A system with an exponentially larger number of dynamic actors demands significantly more computational effort for dynamic analysis. Hence, it is crucial to manage the complexity of future power systems to enable dynamic stability assessments and ensure the safe operation of power grids. 
    
    Model order reduction techniques, such as balanced truncation, are commonly used to simplify parts of power systems \cite{ghosh_balanced_2013}. Transient simulations can be sped up significantly by replacing sub-systems with simpler models. However, ensuring that the reduced model reliably represents the entire system is challenging due to the non-linear and networked nature of power grids. This uncertainty can be problematic: if the reduced sub-system does not accurately reflect the entire system's behavior during transient simulations, it might trigger false alarms when the system is stable or, worse, overlook critical states. Therefore, it is crucial to quantify the accuracy of reduced models.
    
    Recently, the "Probabilistic Behavioral Tuning" (ProBeTune) framework was introduced to address this issue \cite{hellmann_probabilistic_2023}. ProBeTune provides probabilistic distance measures to specify the distance between reduced and full models. Realistic power grids face many possible scenarios where the system is perturbed from its stable operation. The probabilistic approach allows us to explicitly model the scenarios and quantify how well the reduced model represents the full model under these scenarios. 

    In this paper, we apply ProBeTune to the Nordic5 (N5) test case, a model representing the Nordic power grid, which is characterized by complex nodal dynamics and a high share of renewable energy sources (RES) \cite{bjork_dynamic_2022}. This study aims to simplify the dynamics of the Nordic region to a single swing equation at the grid connection point to the Central European (CE) system. The swing equation with a range of permissible parameters specifies the aggregate behavior. Such specifications are well suited to encode the stabilizing behavior of the aggregate subsystem towards the overall system. Achieving this aggregation would allow the CE system to treat the Nordic region as a single, simplified actor, thereby reducing complexity. 
    
    ProBeTune is introduced in detail in section \ref{sec:probetune}. Information on the modeling of the N5 can be found in section \ref{sec:modeling}. The different control designs and scenarios examined in this work are detailed in sections \ref{sec:controllers} and \ref{sec:scenarios}, respectively. We compare the performance of ProBeTune to an analytic baseline, which we introduce in section \ref{sec:baseline}. The main findings of our study are detailed in Section \ref{sec:results}, where we demonstrate that the aggregated behavior of the N5 system can be tuned to behave like a swing equation. This shows that sophisticated grids, subject to unknown disturbances, can be tuned to simple specifications. The results presented in this paper lay the groundwork for future research aimed at aggregating and optimizing the dynamics of power grids using fully differentiable models. 

\section{Methods}
    \subsection{Nordic5}
    \label{sec:modeling}
\begin{figure}[H]
        \centering
        \includegraphics[width=0.6\columnwidth]{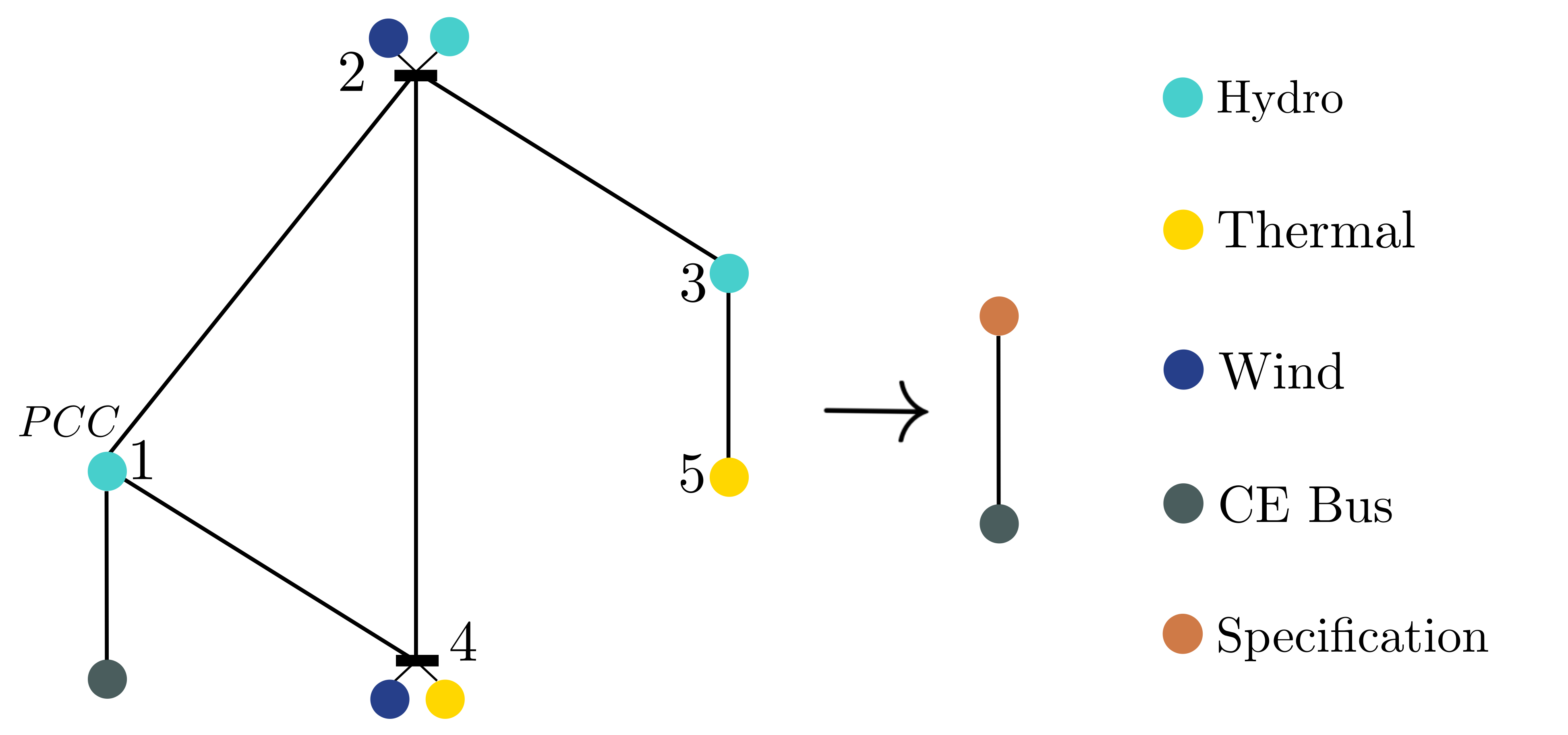}
        \caption{Network structure of the N5 system. Each bus consists of a load, a controller, and additional energy sources. Bus 1 connects the N5 to central Europe via an HVDV link. The goal is to reduce the entire N5 system to the specification shown in orange.
        \label{fig:network}}
    \end{figure}

    The Nordic5 (N5) test case was introduced in \cite{bjork_dynamic_2022} and is an ideal candidate for studying with ProBeTune, as it exhibits intriguing dynamic phenomena. The network structure of the N5 and the desired specification are illustrated in Fig.~\ref{fig:network}. The N5 is densely coupled internally but has only a single connection to Central Europe (CE). Each bus in the N5 features a load and at least one energy source, as indicated in the figure. A third-order machine models the CE bus \cite{machowski_power_2008, schmietendorf_self-organized_2014}, with an additional load at the bus.

    Each bus in the N5 system contains a machine whose shaft sets the frequency dynamics. We have added proportional control $D_i \omega_i$ to each shaft as the baseline form of control. This results in the following frequency dynamics $\omega_i$ for bus $i$:
    \begin{align}
        \dot{\omega_i} &= \frac{1}{2H_i}(P_{m,i} - P_{e,i} - D_i \omega_i) \\
    \end{align}
    where $H_i$ is the inertia constant, $ P_{m, i}$, and $ P_{e, i}$ are the mechanical and electrical power, respectively, that are given by the corresponding differential equations of the machine models, governors, and exciters. 

    Following \cite{bjork_dynamic_2022}, the thermal machines are modeled as sixth-order machines, with an IEEE Type 1 voltage regulator as the exciter, a fourth-order controller as the power system stabilizer (PSS), and a shaft. The hydro machines are represented by a fifth-order machine model, a shaft, and a simplified version of the non-linear governor model introduced in \cite{mover_hydraulic_1992}. The block diagrams depicting these models are provided in the appendix \ref{sec:block}.

    The governor's power production initially undershoots when the reference power $P_{ref}$ is increased \cite{kundur_power_2012}, which could lead to dangerously low frequencies and the triggering of cascading failures. To overcome this issue, the authors of \cite{bjork_dynamic_2022} have equipped the governor with a frequency containment reserve (FCR) controller. The different FCR-control designs are introduced in section \ref{sec:controllers}. The wind turbines are described by the model introduced in \cite{bjork_variable-speed_2022} that includes a grid-following inverter and an additional FCR controller. 

    \subsection{Probabilistic Behavioral Distances}
    \label{sec:probetune}
    This section will introduce the "Probabilistic Behavioral Tuning" (ProBeTune) framework, a probabilistic aggregation technique based on a behavioral approach.

    The behavioral approach defines systems by their behaviors, the sets of inputs, and respective outputs that arise in response to these inputs. This approach focuses on the observable behavior of the system rather than its internal structure. By defining systems through their behaviors, we can directly compare the dynamic response of different systems. This is particularly useful when dealing with complex, interconnected systems such as power grids. The article \cite{willems_behavioral_2007} provides an excellent introduction to the behavioral approach for dynamical systems. Power grids are subject to various sources of uncertainty, such as fluctuating demand, variable renewable energy supply, and unexpected disturbances. A probabilistic approach allows us to model and account for these complex uncertainties explicitly.
    
    In \cite{hellmann_probabilistic_2023}, the authors introduce the concept of probabilistic distance measures for non-linear systems with stochastic inputs. This distance measures how far the behavior of a system is from the behavior of an idealized, reduced specification. The specification is defined as a set of desirable, simple dynamical equations parameterized by a set of parameters $q$. 
    
    For the N5 system, the specification is chosen as the swing equation, representing the idealized behavior in response to a power imbalance. Being close to this idealized behavior indicates that the internal complexities have been effectively hidden from the central European system. Using ProBeTune, we aim to optimize the \emph{controllable} parameters $p$ of the N5 system so that its dynamics closely resemble those described by a single swing equation. In section \ref{sec:controllers}, we describe which parameters are fixed and controllable.
    
    For the N5 system, as seen from the CE grid, a natural choice for the output is the frequency $\omega(t)$ visible at the connection point between the grids. For the N5 system, the input from the CE grid is the current flowing on the transmission line between them. The output metric $o$ is based on the $L_2$-norm between the frequency of the system at the Point of Common Coupling (PCC), $\omega_{pcc}$, and the frequency of the specification, $\omega_{\mathrm{spec}}$. The output metric is this $L_2$ norm averaged over all scenarios $N$ and includes possibly scenario-dependent specification parameters $q_k$:
    \begin{align}
        o(p, q) = \frac{1}{N} \sum_{k=1}^{N} \sum_{t} \left(|\omega_{pcc}^k(p, t) - \omega_{\mathrm{spec}}^k(q_k, t)|\right)^2, \label{eq:loss}
    \end{align}
    where $k$ and $t$ run over the $N$ scenarios and 5000 uniformly distributed time points in the time series, respectively. This metric requires an ensemble of scenarios sampled from a probability distribution $\rho$. It is crucial to choose these ensembles carefully, as the distances can only be guaranteed for the probability distribution from which the ensembles have been drawn. The considered scenarios are discussed in section \ref{sec:scenarios}.
    
    The behavioral distance, as defined in \cite{hellmann_probabilistic_2023}, is given by the minimum of the output metric \eqref{eq:loss} with respect to $q$:
    \begin{align}
        d^{\rho} = \min_{q} o(p, q), \label{eq:behavioral_distance}
    \end{align}
    which means that \emph{only} the parameters of the specification are optimized. This distance can be used to validate how close the system and specification are at different steps of the tuning pipeline. We refer to the distance before any additional N5 parameters $p$ tuning as the initial distance $d^{\rho}_{init}$.
    
    In a second step, the controllable parameters $p$ of the N5 and the specification $q$ are optimized \emph{jointly} such that $\omega_{pcc}(t)$ and $\omega_{spec}(t)$ come as close to each other as possible. To find the set of optimal parameters $\left( p_{opt}, q_{opt} \right)$, the authors of \cite{hellmann_probabilistic_2023} formulate a joint-optimization problem: 
    \begin{align}
        \left( p_{opt}, q_{opt} \right) = \arg \min_{p} \min_{q} o(p, q).
    \end{align}
    With this optimal set of parameters, the behavioral distance after tuning $d^{\rho}_{end}$ is calculated to validate how close the system and specification are. To verify that no over-fitting occurred, $N$ new samples are drawn, and the resampled behavioral distance $d^{\rho}_{re}$ is estimated. If the distance $d^{\rho}_{re}$ does not increase significantly from $d^{\rho}_{end}$, we can be sure that the specification does not only memorize the training samples but adequately represents the system behavior. Hence, we can be confident that the specification accurately captures the dynamic behavior of the N5. All steps of the tuning pipeline are summarized in Table \ref{tab:tuning_steps}. 
    \begin{table}[H]
        \centering
        \begin{tabular}{|l|}
            \hline
            \textbf{Tuning Pipeline Steps}                      \\ \hline
            Draw $N$ random samples from $\rho$                 \\ \hline
            Calculate baseline $o_{base}$                       \\ \hline
            Estimate the initial distance $d^{\rho}_{init}$     \\ \hline
            Tune system and specification to each other         \\ \hline
            Estimate $d^{\rho}_{end}$ after tuning              \\ \hline
            Draw $N$ new samples and calculate $d^{\rho}_{re}$  \\ \hline
        \end{tabular}
        \caption{Steps of the tuning pipeline used in this paper.}
        \label{tab:tuning_steps}
    \end{table}
    The approach has been implemented numerically as calculating the behavioral distance and the optimal parameters is typically analytically intractable. In principle, traditional optimization techniques, such as grid search, can be used to calculate the distance and determine the optimal parameters. However, the resulting computation times are unfeasible for practical applications. To perform these optimizations within reasonable times, gradient-descent methods and auto-differentiation are required. In the literature, ProBeTune has not been applied to realistic power grids, only to conceptual oscillator networks \cite{hellmann_probabilistic_2023}, due to the lack of a fully differentiable power grid model. A major contribution of this work is demonstrating that building and optimizing a fully differentiable complex grid model is feasible. In the appendix \ref{sec:implementation}, we highlight the software and computational methods we have employed to achieve such a fully differentiable system.

    Recently, several papers have been published that address the aggregation of behaviors to achieve desirable dynamics. Among them is \cite{bjork_dynamic_2022}, which introduced the N5 test case. Particularly noteworthy are the publications by Häberle et al. \cite{haberle_control_2022, haberle_grid-forming_2022}, which aim to achieve desired multi-input multi-output behavior. The authors achieve each device's desired global and local behavior and focus on providing ancillary services. In contrast, our focus is on aggregation and thus only specifies a global specification. Notably, in \cite{haberle_control_2022}, only linear parameter-varying systems, a class of non-linear systems that can be expressed as linear systems with state-dependent parameters, can be considered the specification. Our approach, however, allows for utilizing fully non-linear specifications. This is desirable as non-linearities characterize the dynamics of power grids and must be included in an appropriate reduced model for transient simulations.

   \subsection{Specification} 
    In the following, we introduce the specification employed as the reduced model for the N5. The specification is a single swing equation with controlled power generation $P_{m}$. The swing equation is the idealized behavior of a power plant or region to a power imbalance. Being close to this idealized dynamic behavior means that the CE system can treat the Nordic region as the simpler model that facilitates dynamic simulations. The dynamics for the specification are given by:
    \begin{align}
        \dot{\omega} &= \frac{1}{2H}(P_{ref} - P_{e} - D\omega) \nonumber\\
        P_{ref} &= P_{fix} + u(\omega) \label{eq:spec} 
    \end{align}
    where $u(\omega)$ is the control input and $P_{fix}$ is the power consumption of the load in the operation point. We use the same control input $u$ in the specification and before the governors in the N5. In addition to the control parameters, we also allow the inertia constant $H$ of the specification to be tuned.
        
    \subsubsection{Controllers}
    \label{sec:controllers}
    In this section, we introduce the tunable controllers in the N5, especially the FCR controllers employed by the hydro-governor and wind turbines. We have fixed all model parameters of the N5 to the values given in the literature except those of the FCR controller and the proportional control, which we assumed to be adaptable. These parameters are optimized to tune the N5 to behave as the specification. A summary of the controllers and the controllable parameters for the system $p$ and specification $q$ is provided in Table \ref{tab:controller_paras} at the end of the section.
    
    We employ proportional control at the shaft as the baseline and progress towards more elaborate control schemes, such as leaky integral controllers \cite{weitenberg_robust_2019}. Using only proportional control results in a deviation between the asymptotic and nominal frequency when the power is changed. This deviation arises from the power mismatch that cannot be counterbalanced, resulting in undesirable behavior. Despite this, proportional control remains of interest as the swing equation with a proportional term is analytically well understood \cite{dorfler_synchronization_2012}.    

    The FCR-controller, which has been included in the model to prevent undershoots, measures the local frequency and calculates the reference power $P_{ref}$ as follows:
    \begin{align}
        P_{ref,i} &= P_{fix,i} + u_i(\omega_i) \label{eq:FCR}
    \end{align}
    where $P_{fix,i}$ is the power bus $i$ generates in the operation point and $u_i$ is the control input of the FCR-controller at bus $i$. The additional FCR control aims to adjust the power production of all generators to restore power balance and achieve the nominal frequency $\omega_0$. 
    
    First, we added integral control to the FCR controller given in equation \eqref{eq:FCR}, which is given by the following control input:
    \begin{align}
            u_i &= - K_i \int \omega_i dt , \label{eq:I-controller}
    \end{align}
    where $K_i$ is the integral gain at bus $i$. We call the combination of proportional and integral controllers PI-controllers, which aligns with the traditional literature.
    
    Additionally, we have implemented leaky integral control, which was introduced in \cite{weitenberg_robust_2019}. The leaky integral controller is a fully decentralized frequency restoration controller. The authors of \cite{weitenberg_robust_2019} have shown that the leaky integral controller accomplishes a trade-off between performance and robustness and between asymptotic disturbance rejection and transient convergence rate by tuning the control parameters. The control input $u_i$ of the leaky integral controller at bus $i$ is given by:
    \begin{align}
            u_i&= - y_i\\
            \frac{dy_i}{dt} &= \frac{1}{T_i} (\omega_i - G_i y_i). \label{eq:LI-controller}
    \end{align}
    where $T_i$ and $K_i$ are the time and gain constant at bus $i$ respectively. The leaky integral controller, in combination with a proportional term, is referred to as a PLI controller.
    \begin{table}[H]
        \centering
        \begin{tabular}{|l|l|l|}
            \hline
            Controller                       & Parameters System   & Parameters Specification \\ \hline
            P (no additional FCR-Control)    & $D_i$               & $D, H$                   \\ \hline
            PI                               & $D_i, K_i$          & $D, K, H$                \\ \hline
            PLI                              & $D_i, G_i, T_i$     & $D, G, T, H$             \\ \hline
        \end{tabular}
        \caption{Summary of the tuneable parameters for the system and the specification.}
        \label{tab:controller_paras}
    \end{table}
    
\subsection{Scenarios}
    \label{sec:scenarios}
    The probabilistic distances introduced above require an ensemble of scenarios. It is crucial to choose these ensembles carefully as the distances obtained can only be guaranteed for the probability distribution  $\rho$ the scenarios have been drawn from. First, we consider a smooth quasi-random model similar to that used in the ProBeTune paper \cite{hellmann_probabilistic_2023} to induce a complex response in the system. Second, we also use a proper stochastic process for demand fluctuations, using the model introduced in \cite{anvari_data-driven_2022}, which provides a more realistic picture.
    
    The demand of the CE bus is given by $P_d = P_{load} + P_{fluc}(t)$, where $P_{load}$ is the power the bus consumes in the steady state and $P_{fluc}(t)$ is the fluctuation.
    
\begin{figure}[H]
        \centering
        \begin{subfigure}{.5\textwidth}
                \centering
                \includegraphics[width=0.99\columnwidth]{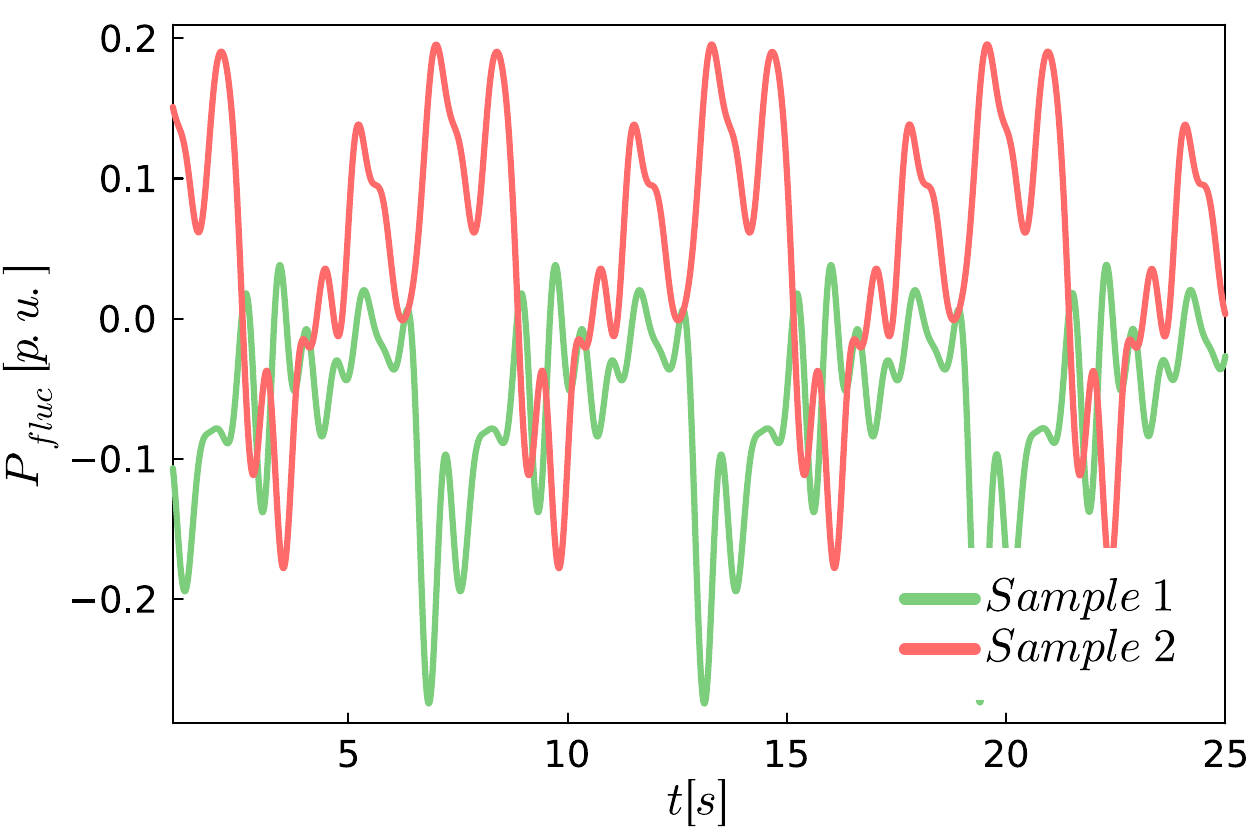}
                \caption{Random modes}
                \label{fig:demand_flucs_mode}
        \end{subfigure}%
        \begin{subfigure}{.5\textwidth}
                \centering
                \includegraphics[width=0.99\columnwidth]{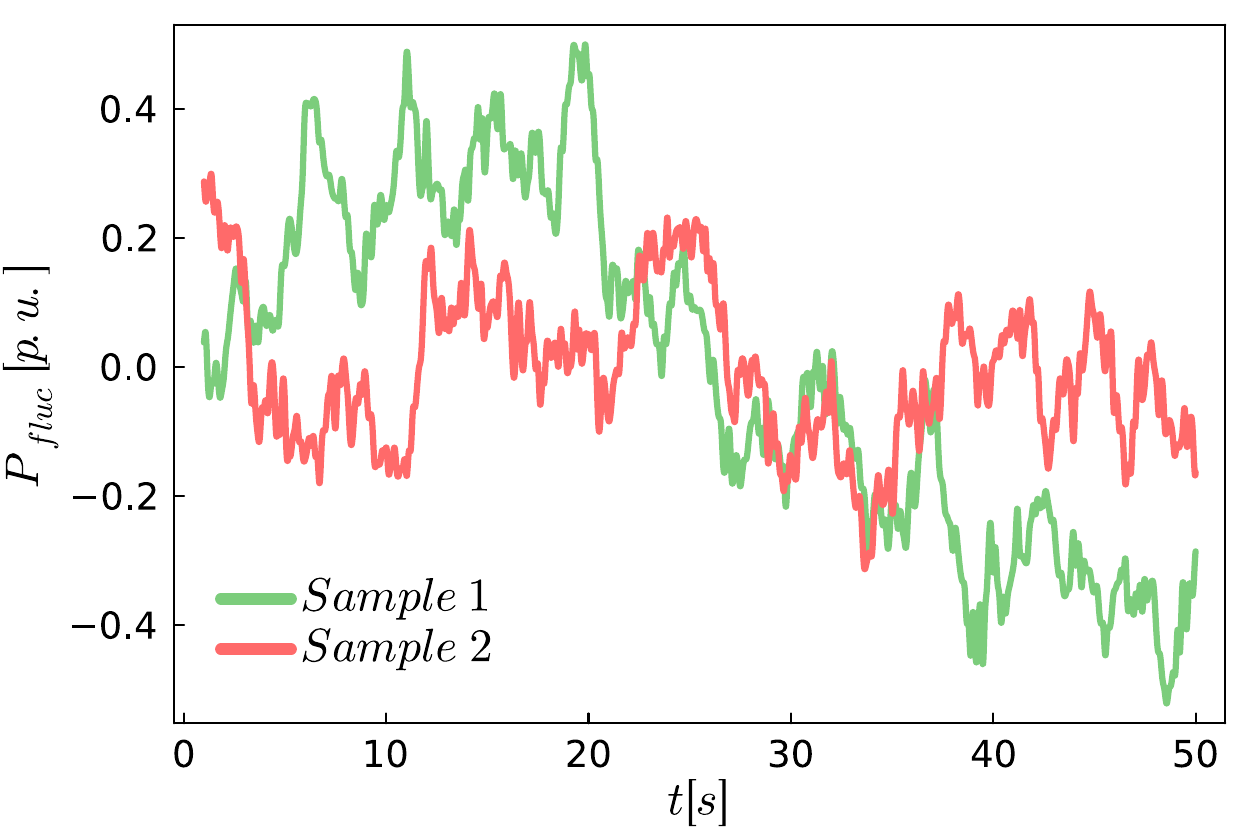}
                \caption{Demand fluctuations}
                \label{fig:real_demand_flucs}
        \end{subfigure}
        \caption{Two realizations of the random modes and the demand fluctuations.}
        \label{fig:}
    \end{figure}

    The fluctuating time series is generated by adding up modes with random amplitudes $A_n$, and phase shifts $\phi_n$ similarly to the set-up in the inputs in \cite{hellmann_probabilistic_2023}. The demand fluctuation is then given by equation \eqref{eq:random_modes}:
    \begin{align}
        P_{fluc}(t) = \sum_{n = 1}^{N_{freq}} A_n \cos(t \cdot n + \phi_n). \label{eq:random_modes}
    \end{align}
    We always use $N_{freq} = 10$ different modes in this example. Figure \ref{fig:demand_flucs_mode} shows two possible realizations of this process.

    We use the demand model introduced in \cite{anvari_data-driven_2022} for realistic demand fluctuations. The authors of the study propose a methodology for extracting both the average demand profiles $P_{trend}(t)$ and the demand fluctuations $P_{fluc}(t)$ from demand time series data. They introduce a stochastic model to capture real-world demand fluctuations that are asymmetric and heavy-tailed. In this work, we only focus on the fluctuation as we only study short-term dynamics. To describe the fluctuations, Ornstein-Uhlenbeck processes are used:
    \begin{align}
        dx_i(t) = -\gamma x_i(t) dt + \epsilon dW_i
    \end{align}
    where $\gamma$ is the damping coefficient, $\epsilon$ is the noise amplitude of the Wiener processes $W$. These processes are used to define the power fluctuation $P_{fluc}(t)$:
    \begin{align}
        P_{fluc}(t) = \sqrt{\sum_{i=1}^J x_i(t)^2} + \mu_{MB} \label{eq:real_demand_fluc}
    \end{align}
    where $\mu_{MB}$ is the observed shift from zero and $J$ is the number of independent Ornstein-Uhlenbeck processes. As suggested by the authors, we use $J=3$ different processes. We used the coefficients $\gamma, \epsilon, \mu_{MB}$ which have been extracted from the NOVAREF data set \cite{m_lange_novaref_2016} that consists of high-resolution demand profiles for 12 German households. Extracting the coefficients from a data set with more consumers would be desirable. Still, to our knowledge, there is no publicly available data set with a sufficient time resolution. Figure \ref{fig:real_demand_flucs} shows two illustrative, realistic demand time series. 
    
\section{Results}
\label{sec:results}

\subsection{Analytic Baseline}
    \label{sec:baseline}
    To assess the performance of the tuned system, we define a baseline derived from analytic considerations. The goal is to find a relation between the parameters of the system and the specification, which brings both dynamics as close to each other as possible. This is an intricate task as the dynamics of the N5 are non-linear and multi-dimensional. The calculations only consider the asymptotic state of the system, as including the transient behaviors is not analytically tractable. In the following, we present the simplifications made for the calculations and the main steps to arrive at the baselines. The full derivation can be found in the appendix \ref{app:baseline}.

    \begin{simplification}
        All buses are modeled as swing equations with controlled power generation, as in equation \eqref{eq:spec}. 
    \end{simplification}
    All buses rotate at the nominal, synchronous frequency $\omega_0 = 0$ during normal operation. Due to the proportional term, the frequency asymptotically reaches a synchronous state $\omega^*$ after a power jump. The synchronous state may not equal the nominal frequency $\omega_0$. The asymptotic frequency $\omega^*$ depends on the asymptotic control action $u^*$. Therefore, we find different results for $\omega^*$ for each controller.

    \begin{simplification}
        Wind plants and hydro machines perfectly follow the reference power $P_{ref}$, meaning that $P_{m, i} = P_{ref, i}$.        
    \end{simplification}

    \begin{simplification}
        The control design and the control parameters of the system are homogeneous.
    \end{simplification}
    This means that the control and respective parameters are equal across all buses, e.g., $D_{i} = D_{sys}$ for all buses $i$. The same holds for the other control parameters $K_i, T_i, G_i$. 

    The inertia constants $H_i$, which determine the initial response, are given for the N5 system and can not be controlled as they are physical properties of the machines. The specification inertia has been chosen such that $\sum H_{i, sys} = H_{spec}$, ensuring the same initial frequency response under our simplifications. 
    
    For the P-controller, there is no additional control, i.e., $u = 0$, which results in the following equation for the asymptotic frequency $\omega^*$:
    \begin{align}
        \omega^* = \frac{\Delta P_{total}}{M D_{sys}}
    \end{align}
    where $M$ is the total number of buses, we define the total power mismatch as $\Delta P_{total}$ as the sum of all power mismatches at the buses: $\Delta P_{total} = \sum_i^M P_{ref, i} .- P_{e, i} $. The total power mismatch $\Delta P_{total}$ is the same for the system and specification. Hence, we find $D_{base} = M D_{sys}$ as the baseline.
    
    For the PI controller, it is known that the asymptotic error, in our case the asymptotic frequency $\omega^*$, always reaches zero. We find the following relation for the asymptotic frequency $\omega^*$:
    \begin{align}
        \omega^{*} &= 0 = \frac{\Delta P_{total} - K_{sys} \sum_i \int_{0}^{\infty} \omega_i(t)}{M D_{sys}}, 
    \end{align} 
    thus the baseline integral gain becomes $K_{base} = M K_{sys}$.

    For the PLI controller, we find the asymptotic control action:
    \begin{align}
        u^{*} = \frac{\omega^*}{G_{sys}}.
    \end{align}
    Using the asymptotic control action $u^*$, we find the asymptotic frequency:
    \begin{align}
        \omega^{*} = \frac{\Delta P_{total}}{M D_{sys} - M (1/G_{sys})}.
    \end{align}
    Hence, $G_{base} = G_{sys}/M$ has to be chosen such that the system and specification end up in the same asymptotic state. We can not define a baseline for the time constant $T$ as it neither influences the asymptotic state nor the initial response of the system. 

    In addition to serving as a baseline, the analytic considerations are vital for the optimization. The parameter space is multidimensional, and its landscape is unknown. Thus, it is crucial to start from a good initial guess that is already close to the global minima to avoid excessive computation times or convergence into local minima.
    
\subsection{Numerical Results}
\subsubsection{Random Modes}
    \label{sec:random_modes}
    As a first step, we benchmarked the simulation times of both the specification and the system to show the possible speed-up. The benchmark shows that the specification runs significantly faster than the system across all control designs. A relative speed-up ranging from approximately 6.42 to 6.83 times can be achieved for one sample. More details on the benchmark can be found in the appendix \ref{sec:benchmark} in table \ref{tab:comparison_modes}.

\begin{figure}[H]
        \centering
        \includegraphics[width=0.5\linewidth]{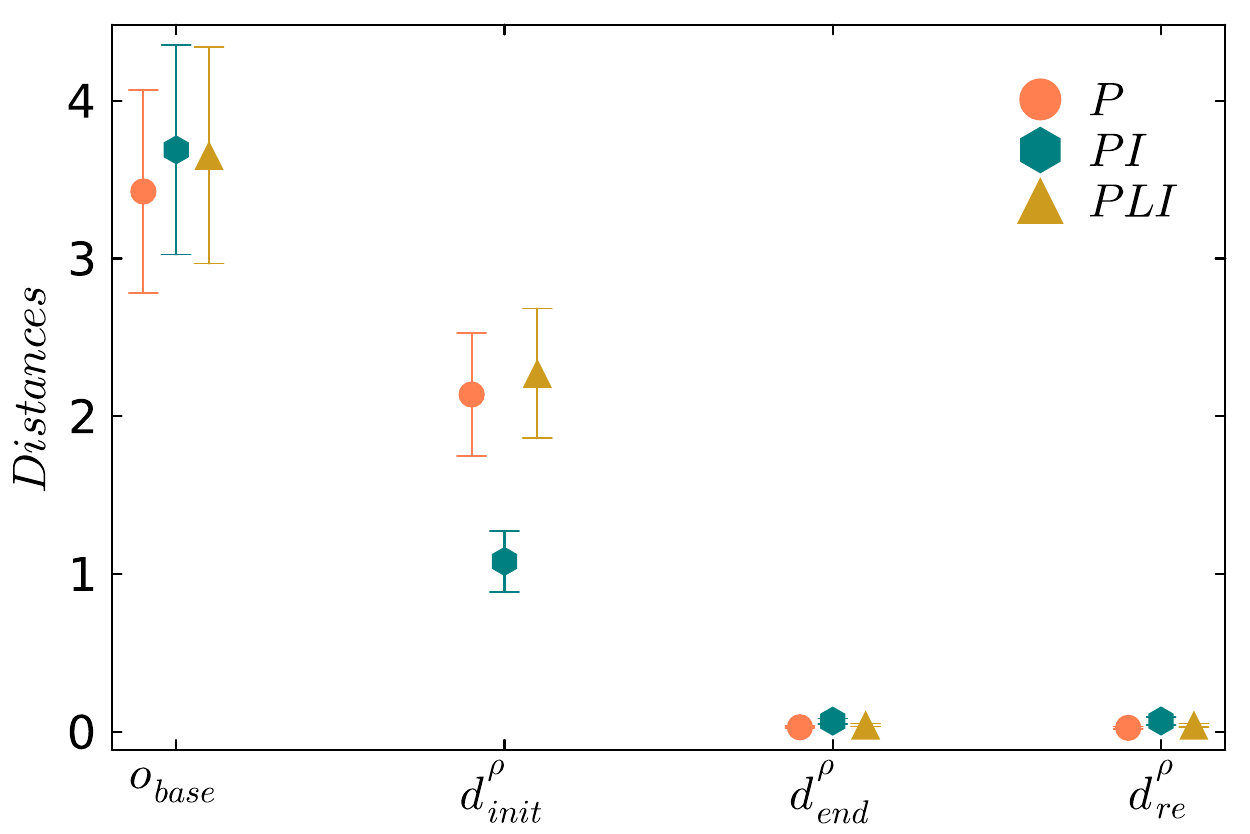}
        \caption{Comparison between the behavioral distances at the different steps in the tuning pipeline for the random mode fluctuations.}
        \label{fig:results_modes}
    \end{figure}

    The following shows the results of applying ProBeTune to the N5 system. Figure \ref{fig:results_modes} illustrates the behavioral distance $d^{\rho}$ between the system and the specification at different stages in the tuning pipeline for the three control strategies. Exact values for the distances are provided in Table \ref{tab:results_modes} in the appendix. Figure \ref{fig:results_modes} shows that the baseline performance $o_{base}$ shows substantial deviations from the desired behavior, with all controllers exhibiting similar deviation levels. 
    
    The initial distances $d^{\rho}_{init}$, calculated using equation \eqref{eq:behavioral_distance}, are significantly better than the baseline performance. This is expected due to the simplifications, especially since only the asymptotic behavior is considered for the baseline.  The initial distances $d^{\rho}_{init}$ exhibit varying degrees of misalignment, with the PI strategy having the least deviation, while the P and PLI controllers show higher deviations. 
    
    After the tuning process, the distances $d^{\rho}{end}$ are significantly reduced across all controllers in the same order of magnitude. As expected, including the controllable parameters of the N5 in the optimization further reduces the behavioral distance. Further, the system parameters $p_{opt}$ after tuning are inhomogeneous, meaning that they are specialized for each bus, unlike for the initial distance $d^{\rho}_{init}$ where homogeneous system control parameters are employed, see section \ref{sec:baseline}.  
    
    Notably, the distances after resampling, $d^{\rho}_{re}$, are identical to $d^{\rho}_{end}$, indicating that no over-fitting has occurred and the system has accurately learned the behavior of the specification. These results demonstrate the effectiveness of the behavioral distance approach in combination with tuning.
    
\begin{figure}[H]
        \centering
        \begin{subfigure}{.33\textwidth}
                \centering
                \includegraphics[width=0.99\columnwidth]{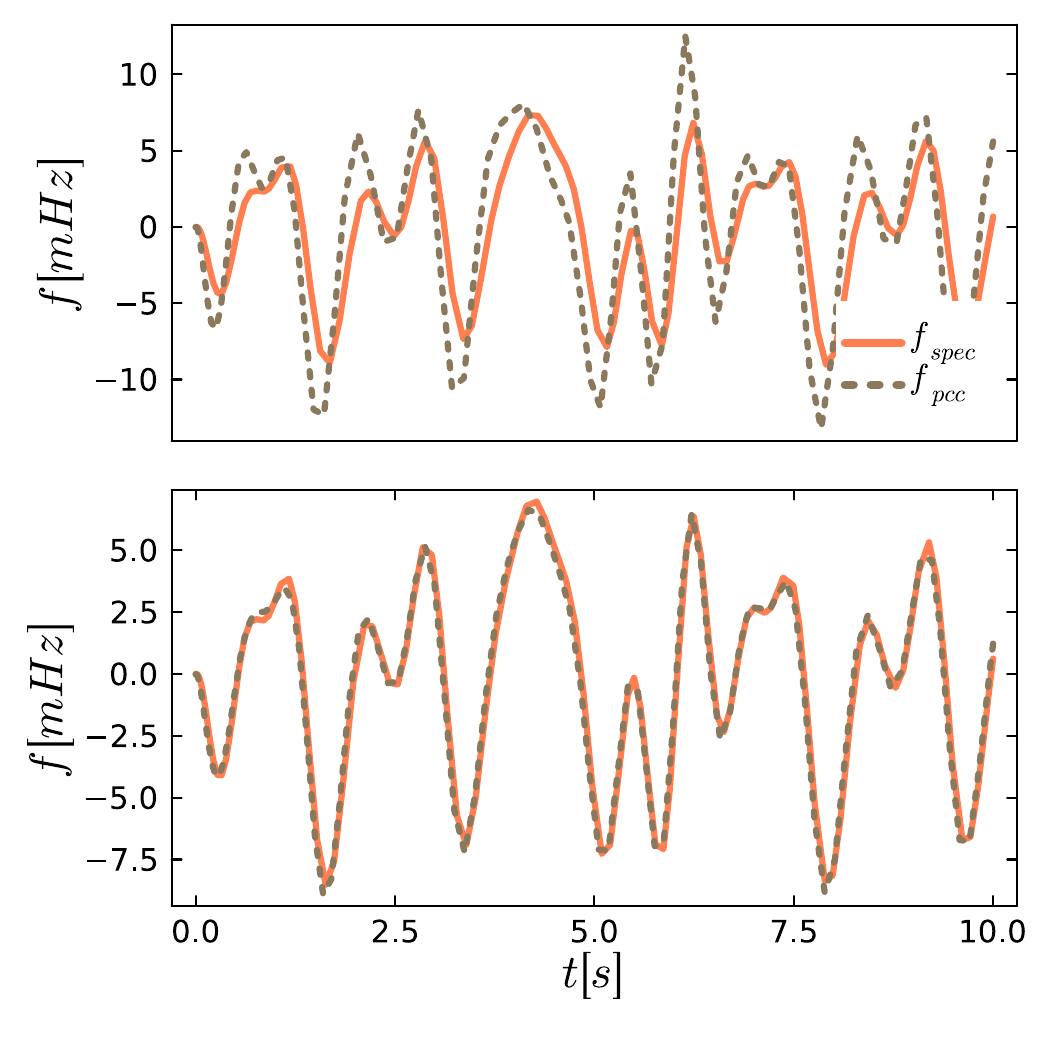}
                \caption{P}
        \end{subfigure}%
        \begin{subfigure}{.33\textwidth}
                \centering
                \includegraphics[width=0.99\columnwidth]{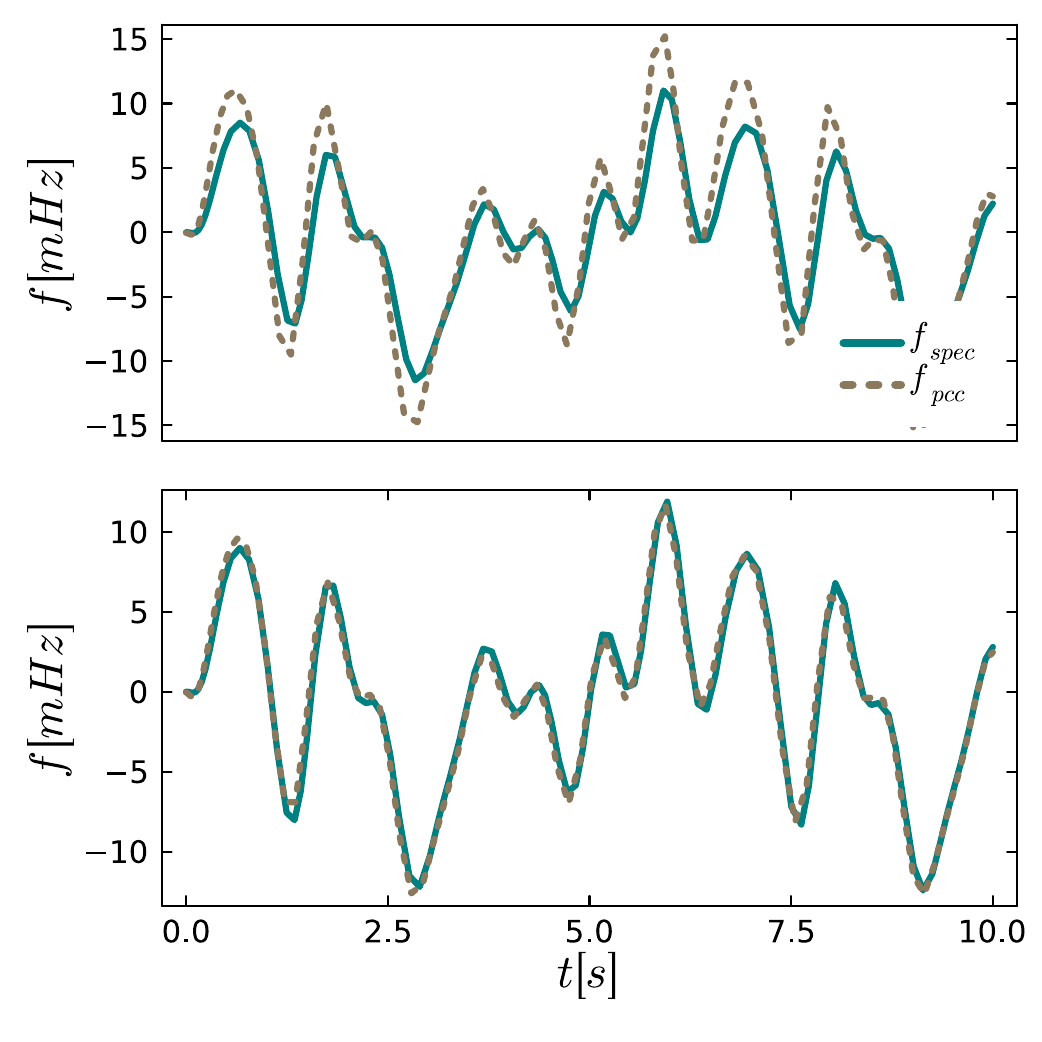}
                \caption{PI}
        \end{subfigure}
        \begin{subfigure}{.33\textwidth}
                \centering
                \includegraphics[width=0.99\columnwidth]{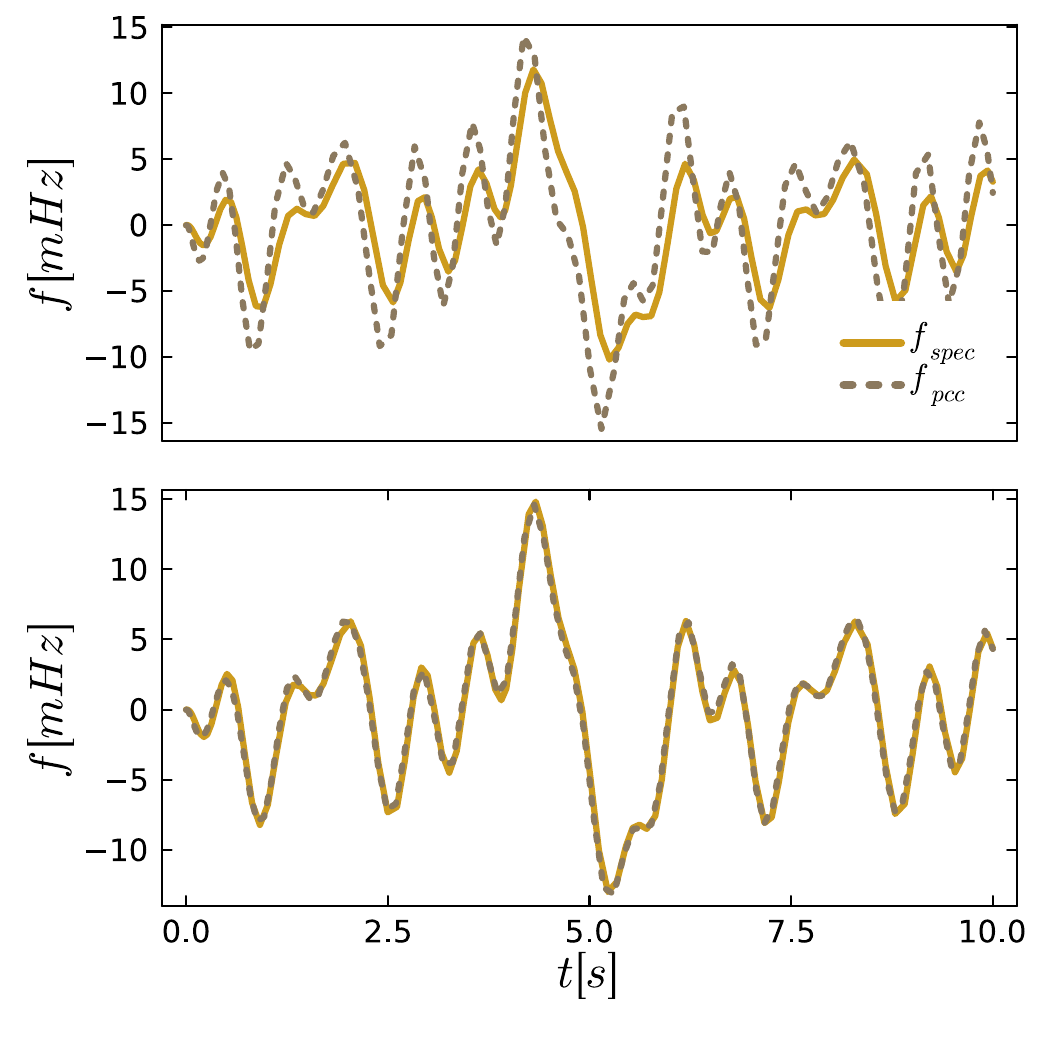}
                \caption{PLI}
        \end{subfigure}
        \caption{Comparison between the system and the specification behavior. The upper figures always show system and specification at the initial distance $d^{\rho}_{init}$, and the lower figure shows them at the distance $d^{\rho}_{end}$ after tuning.}
        \label{fig:comparison_init_end_modes}
    \end{figure}

    To better visualize different levels of behavioral distances, we plot exemplary trajectories before and after the tuning process. Figure \ref{fig:comparison_init_end_modes} shows the frequency $f$ of the specification and the PCC before and after tuning for various scenarios of demand fluctuations. In the pre-tuned state, indicated by the behavioral distance $d_{init}^{\rho}$, the specification and the system exhibit similar reactions to the demand fluctuations, showing the same peaks. However, the shape and amplitudes of these peaks are not well matched. In the tuned state, we observe a close alignment between the shapes and amplitudes of the peaks for all controllers, as anticipated from the small behavioral distances. This further visualizes the effectiveness of the tuning process.
    
\subsubsection{Realistic Demand Fluctuations}
    We have also performed a benchmark for the system and specification for the realistic demand fluctuations. The specification runs significantly faster than the system across all control designs, with a relative speed-up ranging from approximately 18.69 to 22.62 times for a single sample. Further details can be found in the appendix in table \ref{tab:comparison_realistic}.   
    
\begin{figure}[H]
        \centering
        \includegraphics[width=0.5\linewidth]{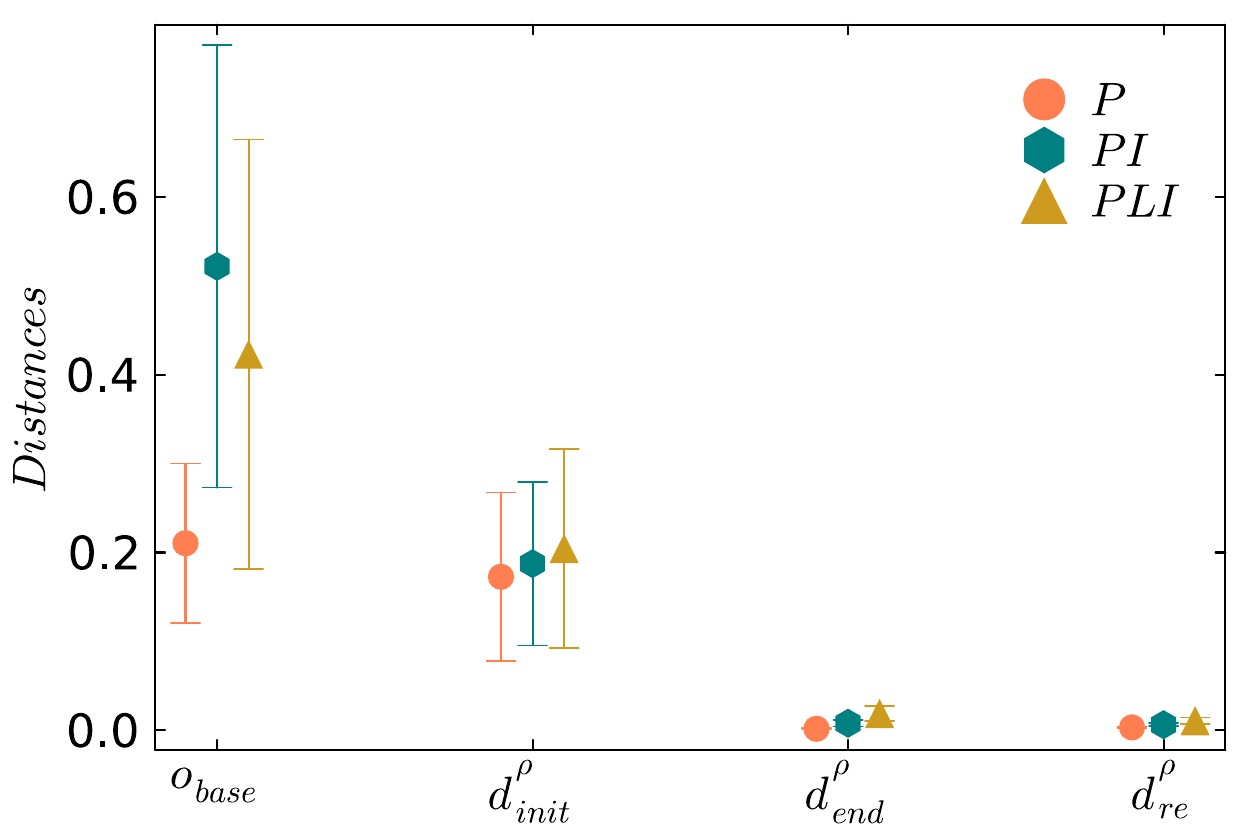}
        \caption{Comparison between the behavioral distances at the different steps in the tuning pipeline for the realistic demand fluctuations.}
        \label{fig:results_realistic}
    \end{figure}

\begin{figure}[H]
        \centering
        \begin{subfigure}{.33\textwidth}
                \centering
                \includegraphics[width=0.99\columnwidth]{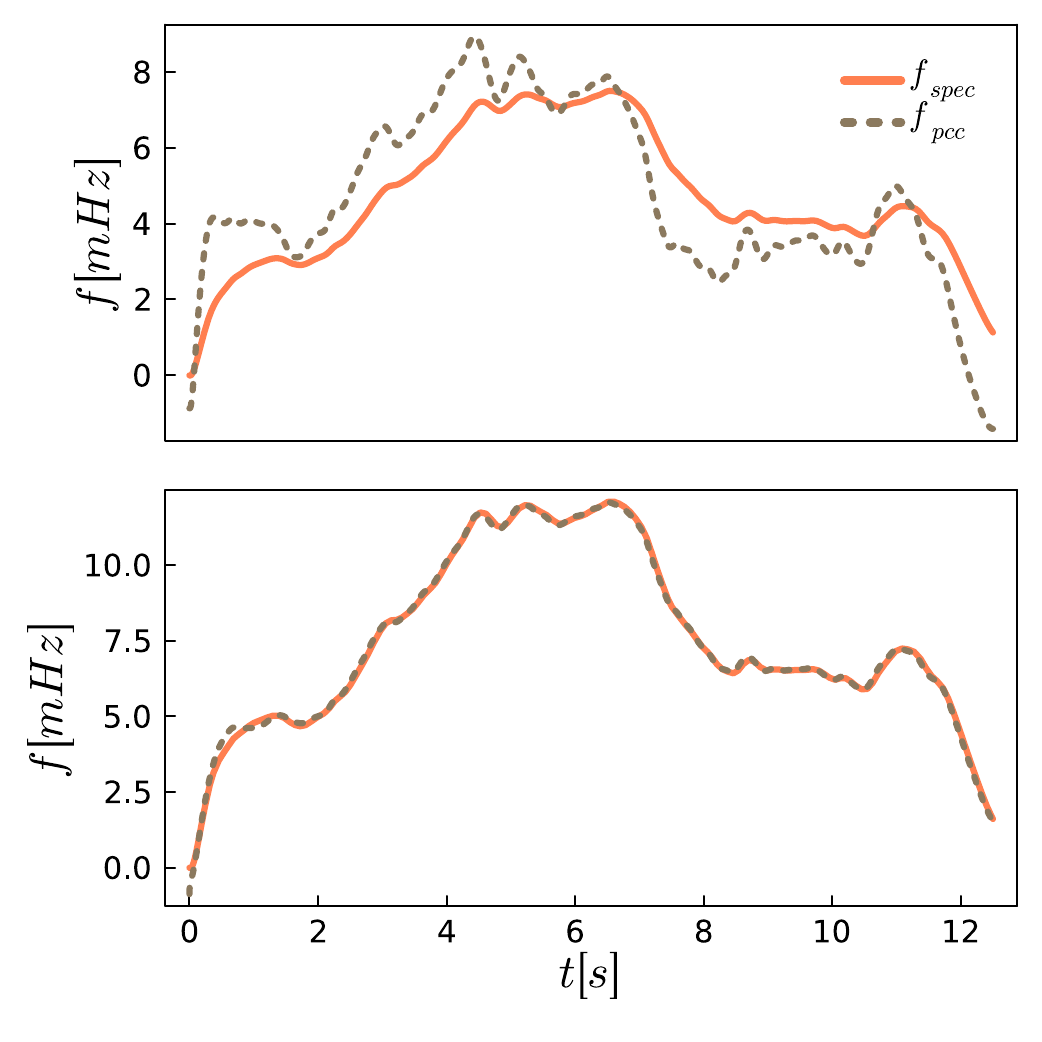}
                \caption{P}
        \end{subfigure}%
        \begin{subfigure}{.33\textwidth}
                \centering
                \includegraphics[width=0.99\columnwidth]{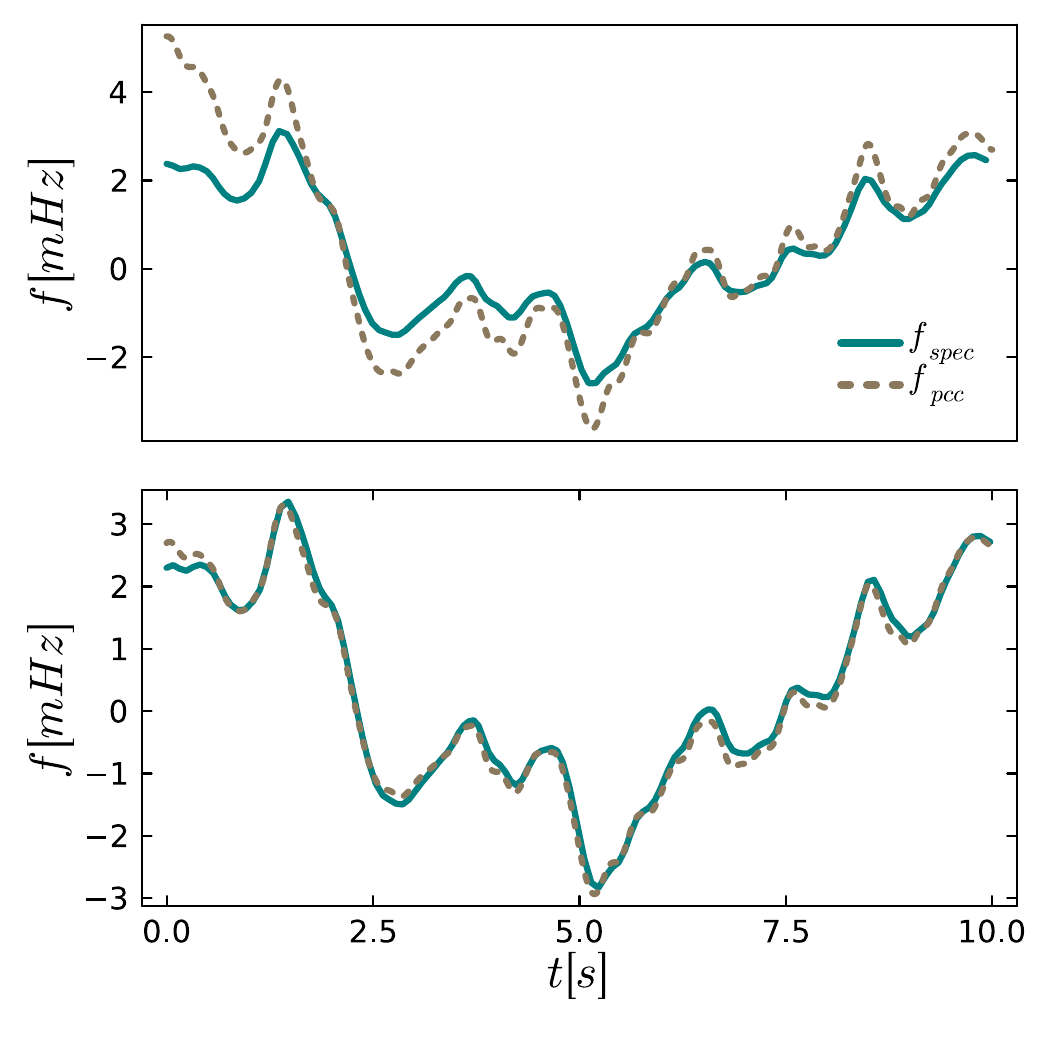}
                \caption{PI}
        \end{subfigure}
        \begin{subfigure}{.33\textwidth}
                \centering
                \includegraphics[width=0.99\columnwidth]{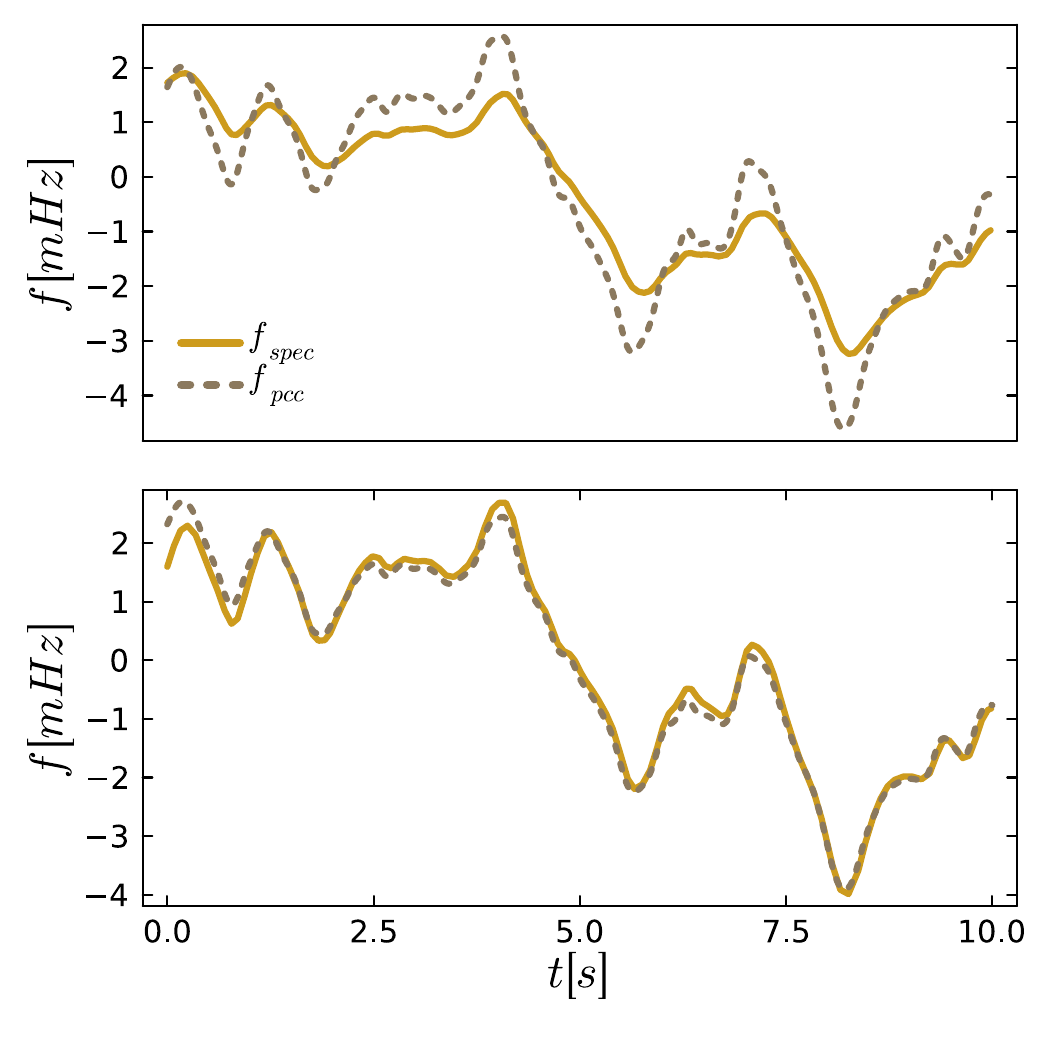}
                \caption{PLI}
        \end{subfigure}
        \caption{Comparison between the system and the specification behavior. The upper figures always show system and specification at the initial distance $d^{\rho}_{init}$ and the lower figure shows them at the distance $d^{\rho}_{end}$ after tuning.}
        \label{fig:comparison_init_end_realistic}
    \end{figure}

    As in the previous section, we analyze the behavioral distances over the different steps in the tuning pipeline. Figure \ref{fig:results_realistic} represents behavioral distance for the realistic demand fluctuations, and the exact values are given in the appendix in Table \ref{tab:results_realistic}. In figure \ref{fig:results_realistic}, we can again see that the baseline $o_{base}$ showcases significant deviations between the system and the desired specification behavior. We can also see that the PI and PLI controllers show more significant deviations than the P controller. As in the previous example, we can see that the initial distances $d_{init}^{\rho}$ are significantly lower than the baselines for all controllers. We can especially see an improvement in the PI controller. After tuning, the distances $d_{end}^{\rho}$ are significantly reduced across all controllers as in the previous example. The resampled distances $d_{re}^{\rho}$ match $d_{end}^{\rho}$, which confirms that no over-fitting has occurred and the system has effectively learned the specified behavior. These results again highlight the success of the tuning processes and the ProBeTune framework overall.
    In Fig.~\ref{fig:comparison_init_end_realistic} we again visualize exemplary trajectories of the untuned and tuned system and specification. A close match between the system and specification is observed for all controllers in the tuned state, consistent with the small behavioral distances. 
    
\section{Discussion}
    In this paper, we have applied the novel ProBeTune concept \cite{hellmann_probabilistic_2023} to realistic power grid dynamics for the first time. ProBeTune is a probabilistic, behavioral strategy to reduce the complexity of networked systems by reducing it to a much simpler specification. The reduction is achieved by introducing the behavioral distance, a measure that specifies the difference between the dynamics of two systems, and then minimizing this distance. The N5 system \cite{bjork_dynamic_2022} has been used as the test case.

    As an initial step, we analytically determined baseline control parameters to align the system's dynamics with the specification. We have shown that behavioral distance applies to power grids and that this distance can be effectively optimized. The behavioral distance has been successfully reduced by orders of magnitude for all experiments. The results of this study show that the swing equation with appropriate controllers and parameter choices is an excellent and efficient model that can aggregate parts of the power grid. This has been demonstrated by using different controllers and various scenarios. Depending on the configuration, a simulation speed-up of 6.42 up to 22.62 times can be achieved by replacing the system with the specification.
    
    These results are particularly interesting for understanding the dynamics of large interconnected systems, which will consist of many microgrids in the future. These lower-level micro-grids are less well understood than transmission systems, and detailed models are scarce, e.g., the demands of households are hard to predict. These lower-level grids will be characterized by high complexity due to the stochastic nature of renewable energy production and demand and the high number of consumers and producers. An aggregation using ProBeTune is helpful so that the individual sub-grids can still be considered in interconnected power grid models. The results presented in this paper lay the foundation for future research in this direction.

\section*{Acknowledgment}
    A. Büttner acknowledges support from the German Academic Scholarship Foundation. The work was in parts supported by DFG Grant Number KU 837/39-2 and BMWK Grant Number 03EI1016A. All authors gratefully acknowledge the European Regional Development Fund (ERDF), the German Federal Ministry of Education and Research, and the Land Brandenburg for supporting this project by providing resources on the high-performance computer system at the Potsdam Institute for Climate Impact Research.

\section*{Data Availability Statement}
    Both the data that supports the findings of this study and the source code used to produce the results are openly available on Zenodo \cite{buttner_probabilistic_2024}.

\printbibliography

\section*{Supporting Information}

    \subsection*{Implementation}
    \label{sec:implementation}
    The key technical challenge in evaluating the probabilistic distance and solving the optimization problem is obtaining fully differentiable, realistic, and fast models of power grids. The following summarizes the computational methods and software used to address this complex joint-optimization problem. We refer interested readers to \cite{buttner_open_2022} for a more detailed discussion of the software stack.

    The entire software stack is implemented using the Julia programming language. One of Julia’s most significant features is its support for "differentiable programming," which enables the efficient and accurate computation of derivatives for arbitrary Julia programs. This capability facilitates using gradient-descent-based optimization methods in conjunction with differential equation solvers. Numerically, the minimization of the behavioral distance \eqref{eq:loss} is performed using a gradient-descent approach with the ADAM optimizer \cite{kingma_adam_2017}.

    Our power system simulations are built on two Julia software packages: \texttt{BlockSystems.jl} and \texttt{NetworkDynamics.jl}. Both packages are designed for highly efficient transient stability simulations of power grids. They allow users to design power systems that are modular and equation-based while maintaining high performance and detail.
    
\subsection*{Block Diagrams}
    \label{sec:block}

    \begin{figure}[H]
        \centering
        \begin{subfigure}{.33\textwidth}
                \centering
                \includegraphics[width=0.99\columnwidth]{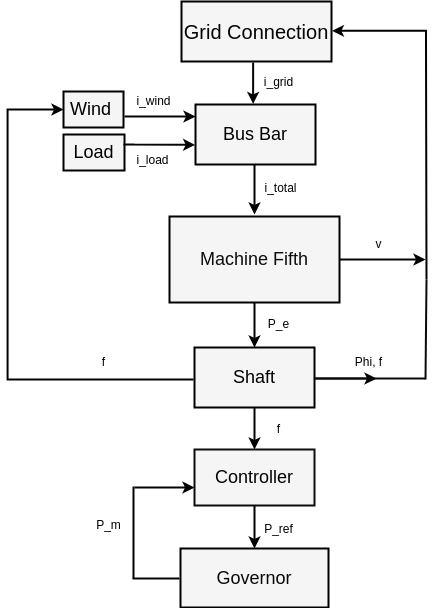}
                \caption{Hydro}
                \label{fig:hydro}
        \end{subfigure}%
        \begin{subfigure}{.33\textwidth}
            \centering
            \includegraphics[width=0.99\columnwidth]{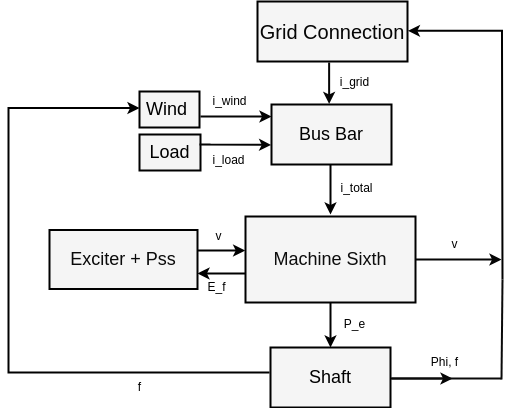}
            \caption{Thermal}
            \label{fig:thermal}
        \end{subfigure}
        \begin{subfigure}{.33\textwidth}
            \centering
            \includegraphics[width=0.4\columnwidth]{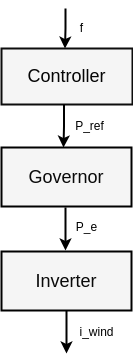}
            \caption{wind}
            \label{fig:wind}
            \end{subfigure}
        \caption{Block Diagrams}
        \label{fig:block_diagrams}
    \end{figure}
    
\subsection*{Analytic Baseline}
    \label{app:baseline}
    Following the simplifications given in \ref{sec:baseline} we derive the analytical baselines. The following system of swing equations describes the approximated system:
    \begin{align}
        \boldsymbol{\dot{\theta}} &= \boldsymbol \omega \\
       2 \bm{H}  \boldsymbol{\dot{\omega}} &= - \bm{D} \boldsymbol{\omega} + \boldsymbol{\Delta P} + \bm{u} \label{eq:vec_swing}
    \end{align}
    where $\boldsymbol{\Delta P} = \bm{P_{fix}} - \bm{P_{e}}$ and $\bm{u}$ is the control input. Multiplying equation \eqref{eq:vec_swing} by the unity vector $\bm{{1}_n^T}$ results in the following relation:
    
    \begin{align}
        \sum_i^M 2 H_i \dot{\omega_i} =& -\sum_i^M D_i \omega_i + \sum_i^M \Delta P_i + \sum_i^M u_i \label{eq:swing_sum}.
    \end{align}
    Where $i$ is the bus index. We define the the total power mismatch $\Delta P_{total}$ as:
    \begin{align}
        \Delta P_{total} = \Delta P_i.
    \end{align}
    Starting again from equation \eqref{eq:swing_sum}, we assume that the asymptotic state is fully synchronized, meaning that $\omega_i = \omega^*$, but not necessarily synchronized at the operating frequency $\omega_0$, which results in the following equation:
    \begin{align}
       0 =& \Delta P_{total} -\omega^{*} \sum_i^M D_i + \sum_i^M u_i^*. \label{eq:u_asymtotic}
    \end{align}
    For the P-controller, there is no additional control, i.e., $u = 0$, which results in the following equation for the asymptotic frequency $\omega^*$:
    \begin{align}
        \omega^* = \frac{\Delta P_{total}}{\sum_i^M D_i} ,
    \end{align}
    where $M$ is the total number of buses, as the total power mismatch $\Delta P_{total}$ is the same for system and specification, we find $D_{base} = \sum_i^M D_{i, sys} = M D_{sys}$ for the baseline.
    
    For the I-controller \eqref{eq:I-controller}, the asymptotic control action $u_i^*$ becomes:
    \begin{align}
        u_i^{*} &= -K_i \int_{0}^{\infty} \omega_i(t)
    \end{align}
    For the PI controller, it is known that the asymptotic error, in our case the asymptotic frequency, always reaches zero. Using these result in equation \eqref{eq:u_asymtotic} results in the following relation for the asymptotic frequency $\omega^*$:
    \begin{align}
        0 &= \Delta P_{total} - \omega^{*} \sum_i^M D_i - \sum_i^M K_i \int_{0}^{\infty} \omega_i(t)  \\
        \omega^{*} &= 0 = \frac{\Delta P_{total} - \sum_i^M K_i \int_{0}^{\infty} \omega_i(t)}{\sum_i^M D_i}.
    \end{align}
    thus the baseline integral gain becomes $K_{base} = M K_{sys}$.
    
    For the leaky integral control \eqref{eq:LI-controller}, we use the separation of variables to solve the first-order differential equation. To find the asymptotic control action $u^*$, we again use the simplification that in the asymptotic state $\omega_i = \omega^*$:
    \begin{align}
        \int \frac{dt}{T_i} &= \int \frac{dy_i}{\omega_i -G_i \cdot y_i} \\
        u_i &= -\frac{1}{G_i} (e^{-t (G_i / T_i)} - \omega) \\
        u_i^* &= \frac{\omega^*}{G_i}.
    \end{align}
    As $e^{-t G_i / T_i}$ goes to zero in the limit of $t \rightarrow \infty$, we can not define a baseline for $T_i$ as it does not influence the asymptotic state nor the initial response of the system. Using the leaky asymptotic control gain and equation \eqref{eq:u_asymtotic}, we find the asymptotic frequency for the system with leaky integral controllers:
    \begin{align}
        0 = \Delta P_{total} - \omega^{*} \sum_i^M D_i + \omega^{*} \sum_i^M \frac{1}{G_i} \\
        \omega^{*} = \frac{\Delta P_{total}}{\sum_i^M D_i - \sum_i^M (1/G_i)}.
    \end{align}
    Meaning that we should choose $1 / G_{base} = \sum_i^M (1 / G_{i, sys}) = M / G_{sys}$ such that the N5 system and specification end up in the same asymptotic state. 

\subsection{Benchmark}
    \label{sec:benchmark}
    This section summarizes the benchmark results between the system and specification. We compare the times for the evaluation of one sample. We have employed \texttt{BenchmarkTools.jl} for the performance tracking. The benchmark has been performed on a Dell Inc. Latitude 7440 with a 13th Gen Intel i7-1365U (12) CPU.
    
    \begin{table}[H]
    \centering
    \begin{tabular}{|l|c|c|c|}
        \hline
        & P & PI & PLI \\
        \hline
        System & \SI{18.865}{ms} $\pm$ \SI{7.045}{ms} & \SI{19.252}{ms} $\pm$ \SI{6.892}{ms} & \SI{20.096}{ms} $\pm$ \SI{5.663}{ms} \\
        \hline
        Specification & \SI{2.939}{ms} $\pm$ \SI{2.881}{ms} & \SI{2.821}{ms} $\pm$ \SI{2.753}{ms} & \SI{3.021}{ms} $\pm$ \SI{2.812}{ms} \\
        \hline
        Relative speed-up & $\approx 6.42$ & $\approx 6.83$ & $\approx 6.65$ \\
        \hline
    \end{tabular}
    \caption{Comparison of the simulation times for the system and specification using random mode fluctuations.}
    \label{tab:comparison_modes}
    \end{table}

    \begin{table}[H]
    \centering
    \begin{tabular}{|l|c|c|c|}
        \hline
        & P & PI & PLI \\
        \hline
        System & \SI{141.028}{ms} $\pm$ \SI{6.931}{ms} & \SI{144.363}{ms} $\pm$ \SI{10.759}{ms} & \SI{193.455}{ms} $\pm$ \SI{6.466}{ms} \\
        \hline
        Specification & \SI{6.660}{ms} $\pm$ \SI{5.239}{ms} & \SI{6.382}{ms} $\pm$ \SI{5.156}{ms} & \SI{10.348}{ms} $\pm$ \SI{5.165}{ms} \\
        \hline
        Relative speed-up & $\approx 21.18$ & $\approx 22.62$ & $\approx 18.69$ \\
        \hline
    \end{tabular}
    \caption{Comparison of the simulation times for the system and specification using the realistic demand fluctuations.}
    \label{tab:comparison_realistic}
\end{table}

\subsection*{Behavioral Distances}
    \begin{table}[H]
        \centering
        \begin{tabular}{|l|l|l|l|}
            \hline
             & P & PI & PLI \\
            \hline
            Baseline $o_{base}$ & 3.425 $\pm$ 0.643 & 3.688 $\pm$ 0.664 & 3.653 $\pm$ 0.686 \\
            \hline
            Initial $d^{\rho}_{init}$ & 2.138 $\pm$ 0.39 & 1.08 $\pm$ 0.194 & 2.272 $\pm$ 0.41 \\
            \hline
            Tuned $d^{\rho}_{end}$ & 0.028 $\pm$ 0.006 & 0.067 $\pm$ 0.016 & 0.043 $\pm$ 0.009 \\
            \hline
            Resampled $d^{\rho}_{init}$ & 0.025 $\pm$ 0.009 & 0.068 $\pm$ 0.025 & 0.042 $\pm$ 0.01 \\
            \hline
        \end{tabular}
        \caption{Behavioral distance $d^{\rho}$ between system and specification at the different steps in the tuning pipeline. The error is given by the standard deviation.}
        \label{tab:results_modes}
    \end{table}

    \begin{table}[H]
        \centering
        \begin{tabular}{|l|l|l|l|}
            \hline
             & P & PI & PLI \\
            \hline
            Baseline $o_{base}$ & 0.21 $\pm$ 0.09 & 0.522 $\pm$ 0.249 & 0.423 $\pm$ 0.242 \\
            \hline
            Initial $d^{\rho}_{init}$ & 0.173 $\pm$ 0.095 & 0.187 $\pm$ 0.092 & 0.205 $\pm$ 0.112 \\
            \hline
            Tuned $d^{\rho}_{end}$ & 0.002 $\pm$ 0.001 & 0.008 $\pm$ 0.004 & 0.019 $\pm$ 0.008 \\
            \hline
            Resampled $d^{\rho}_{init}$ & 0.003 $\pm$ 0.001 & 0.006 $\pm$ 0.002 & 0.011 $\pm$ 0.004 \\
            \hline
        \end{tabular}
        \caption{Behavioral distance $d^{\rho}$ between system and specification at the different steps in the tuning pipeline. The error is given by the standard deviation.}
        \label{tab:results_realistic}
    \end{table}
    
\end{document}